\documentclass[prl,aps,twocolumn,amsmath,amssymb]{revtex4}
\usepackage{graphicx}
\usepackage{dcolumn}
\usepackage{bm}
\begin{document}

\preprint{APS/123-QED}
\title{Single-Electron Effects in a Coupled Dot--Ring System}

\author{L. Meier,$^1$ A. Fuhrer,$^1$ T. Ihn,$^1$ K. Ensslin,$^1$ W. Wegscheider,$^2$ and M. Bichler$^3$}

\affiliation{
$^1$Solid State Physics Laboratory, ETH Z\"urich, 8093 Z\"urich, Switzerland\\
$^2$Institut f\"ur experimentelle und angewandte Physik, Universit\"at Regensburg, Germany\\
$^3$Walter Schottky Institut, Technische Universit\"at M\"unchen, Germany}

\date{\today}

\begin{abstract}
Aharonov--Bohm oscillations are studied in the magnetoconductance of a micron-sized open quantum ring coupled capacitively to a Coulomb-blockaded quantum dot. As the plunger gate of the dot is modulated and tuned through a conductance resonance, the amplitude of the Aharonov - Bohm oscillations in the transconductance of the ring displays a minimum. We demonstrate that the effect is due to a single-electron screening effect, rather than to dephasing. Aharonov--Bohm oscillations in a quantum ring can thus be used for the detection of single charges. 
\end{abstract}


\maketitle

Coupled mesoscopic systems are on top of the agenda for experimentalists trying to achieve charge read-out, controlled interference, and entanglement of electrons in magnetotransport experiments. In general, there are two limiting cases how coupling can be achieved: it can either be realized by pure electrostatic interaction without the overlap of wavefunctions of the systems to be coupled\;\cite{97levinson,97gurvitz,00buttiker,01pohjola,02pilgram}, or by tunnel coupling involving strong wavefunction overlap and limited electrostatic interaction. A number of experiments employed a quantum point contact as a non-invasive probe to detect the charge state of single or double quantum dots by way of electrostatic coupling between the two mesoscopic systems\;\cite{93field,00cooper,02sprinzak,02smith,03gardelis,03elzerman}. 
Furthermore, a quantum point contact has been used as a detector of the charge state of an antidot\;\cite{00kataoka}. However, in mesoscopic systems where the phase coherence of at least one of the two electrostatically coupled partners is important, additional considerations such as dephasing due to quantum measurement have to be taken into account. Examples of such systems are tunnel coupled double dots\;\cite{02smith,03elzerman,03gardelis} and quantum dots in the Kondo regime\;\cite{02wilhelm,02sprinzak}.
These experiments focus on the coupling between quantum dots, or the read-out of a (coupled) quantum dot system via a nearby quantum point contact. 

Our experiments study the coupling of a quantum dot to a ring, in which the occurrence of Aharonov--Bohm (AB) oscillations relies on the phase-coherence of the electrons.
Ring-like geometries allow the observation of the interference of partial waves as a function of magnetic field, i.e., the AB effect\;\cite{59aharonov,87timp,01hansen,03ihn}, and information about electron decoherence can be extracted\;\cite{01hansen,03ihn}. A mesoscopic detector situated nearby is expected to reduce the phase coherent AB oscillation amplitude\;\cite{97levinson}. Such a `which path' experiment has been performed by Buks and coworkers, who coupled a quantum dot embedded in one arm of an AB interferometer electrostatically to a quantum point contact\;\cite{98buks}. 

Here we report on two electrostatically coupled quantum structures, namely, an AB ring and a Coulomb-blockaded quantum dot. The goal of the experiment is to detect the interaction between the two subsystems and to unravel its consequences. We show that single-electron charging in the quantum dot can be detected through a reduced AB oscillation amplitude in the transconductance of the ring. We find that our experimental observations are explained by a single-electron screening effect. 

\begin{figure}[b]
\begin{center}
\includegraphics[width=3.4in]{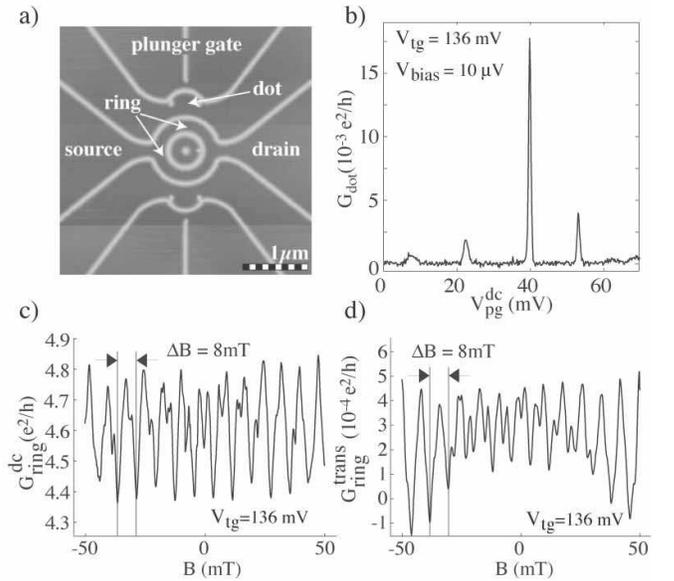}
\caption{(a) Scanning force microscope image of the quantum ring coupled to a quantum dot. The structure was written by AFM lithography.
(b) Coulomb-blockade oscillations in the conduction through the quantum dot as a function of the plunger-gate voltage.
(c) Aharonov--Bohm oscillations in the dc conductance of the ring. The AB period as expected from the ring area is marked by the horizontal arrows. The bias voltage was 40\,$\mu$V, the top-gate voltage $136$\,mV.
(d) Aharonov--Bohm oscillations in the transconductance of the ring. The ac voltage on the plunger gate was modulated at 89 Hz with an amplitude of 1.8 mV.
}\label{fig1}\end{center}\end{figure}

The coupled ring--dot structure was realized on a Ga[Al]As heterostructure with a high-quality two-dimensional electron gas (2DEG) 34\,nm below the surface. The electron density in the 2DEG was $n_\mathrm{s}=5.5\times 10^{15}$\,m$^{-2}$ and the mobility was $\mu=40$\,m$^2$/Vs at 1.7\,K. Ring and dots shown in Fig.\,\ref{fig1}(a) were defined by AFM lithography. Details about the fabrication process can be found in Ref.\,\onlinecite{02fuhrer}. The two-dimensional electron gas is depleted below the oxide lines. Thereby the central AB ring with a diameter of $1\,\mu$m is defined, flanked by quantum dots on both sides. Only one dot is used for the experiments described here. The regions of 2DEG adjacent to the quantum dots are split by an additional oxide line leaving point-contact-like openings. Here, one of these regions is used as the `plunger gate' indicated in the figure. The entire structure was covered by a top gate giving additional tunability. Experiments were carried out in a dilution refrigerator with a base temperature of 40\,mK.

The quantum dot can be tuned into the Coulomb-blockade regime. We determined the electrostatic lever arm of the plunger gate $\alpha_\mathrm{pg}=0.047$ from Coulomb-blockade diamonds\;\cite{04ihn} and the charging energy of the dot is of the order of 0.5\,meV. Pronounced Coulomb-blockade oscillations are observed in the dot conductance $G_\mathrm{dot}^\mathrm{dc}$  [Fig.\,\ref{fig1}(b)]. For the measurements presented below, the circuit containing the dot was lifted by 245\,mV relative to the circuit of the ring. This keeps the ring conductance high while the dot is in the Coulomb-blockade regime.

We express the conductance of the ring with area $A$ in a magnetic field $B$ as
\begin{eqnarray*} G_\mathrm{ring}^\mathrm{dc}(V_\mathrm{pg}^\mathrm{dc},B)&=&G_0^\mathrm{dc}(V_\mathrm{pg}^\mathrm{dc})\\
 & & +G_1^\mathrm{dc}(V_\mathrm{pg}^\mathrm{dc})\cos\left(2\pi \frac{BA}{h/e}\right)+\cdots.
 \end{eqnarray*}
At a constant top-gate voltage $V_\mathrm{tg}=136$\,mV,  the quantum ring structure has a conductance $G_0^\mathrm{dc}= 4.6e^2/h$ that is rather insensitive to small gate-voltage changes. The conductance exhibits pronounced AB oscillations as a function of magnetic field with a fundamental period $\Delta B_\mathrm{AB}=8$\,mT [Fig.\,\ref{fig1} (c)] which corresponds to one flux quantum penetrating the ring area. The relative amplitude of the dc-AB oscillations $G_1^\mathrm{dc}/G_0^\mathrm{dc}$ is independent of the dot conductance on the level of $10^{-3}$.

In order to increase the measurement sensitivity we measure 
the AB oscillations in the transconductance
\begin{eqnarray*} 
G_\mathrm{ring}^\mathrm{trans}&=&\frac{\partial I_\mathrm{ring}^\mathrm{dc}}{\partial V_\mathrm{pg}^\mathrm{dc}}=\frac{\partial G_\mathrm{ring}^\mathrm{dc}}{\partial V_\mathrm{pg}^\mathrm{dc}}V_\mathrm{ring}^\mathrm{bias}\\&=&G_0^\mathrm{trans}+G_1^\mathrm{trans}\cos\left(2\pi \frac{BA}{h/e}\right)+\cdots 
 \end{eqnarray*} which is the derivative of the ring current relative to the voltage applied to the plunger gate.
This quantity is measured with lock-in techniques by applying a dc bias voltage $V_\mathrm{ring}^\mathrm{bias}=40\mu\mathrm{V}$ between source and drain of the ring and modulating the plunger gate with $V_\mathrm{pg}^\mathrm{ac}=300\mu\mathrm{V}$ at a frequency of 89\,Hz. The transconductance, i.e. the modulation of the current $I_{ring}^{dc}$ through the ring, is then detected at the same frequency. All ring conductance data presented in this paper, $G_\mathrm{ring}^\mathrm{dc}$ , are taken in the linear response regime. A similar arrangement to measure a transconductance was used in Ref.\;\onlinecite{98buks}. Figure\,\ref{fig1}(d) shows the resulting AB oscillations in the same magnetic field range as the dc-AB effect in Figure\,\ref{fig1}(c). The fundamental period $h/e$ as well as higher harmonics are visible.

\begin{figure}[t]
\begin{center}
\includegraphics[width=3.4in]{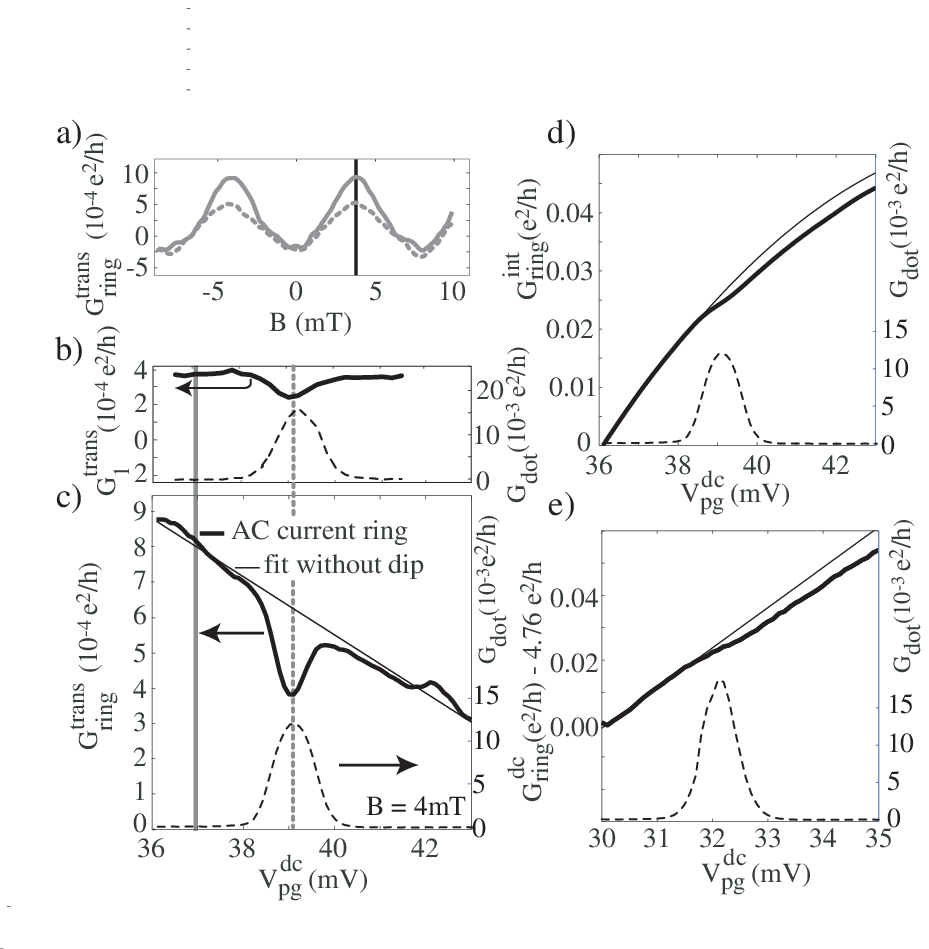}
\caption{(a)~AB oscillations in the transconductance $G_\mathrm{ring}^\mathrm{trans}$ for two different gate voltages $ V_\mathrm{pg}^\mathrm{dc}$= 37 mV and 39 mV, respectively [indicated in (b) by the vertical lines]. (b)~
    AB oscillation amplitude $G_1^\mathrm{trans}$ (full line) obtained from the magnetic field range shown in (a), and dot conductance (dashed line) as a function of plunger-gate voltage. The AB amplitude is
    reduced by about 30\% when the dot shows a conductance peak. 
(c)~Dot conductance peak (dashed line) and ring transconductance $G_\mathrm{ring}^\mathrm{trans}(V_\mathrm{pg}^\mathrm{dc}, $B=4$\,$mT) (full line). The data is taken at a fixed magnetic field of 4\,mT where the effect is largest [see (a)]. The straight line represents the expected behavior if there was no conductance peak in the dot at this gate voltage.
    (d)~Integrated transconductance signal from (c) (full line), and dot conductance (dashed line). The straight line is again the extrapolation without charging effects. $G_\mathrm{ring}^\mathrm{dc}(V_\mathrm{pg}^\mathrm{dc})$
     is expected to exhibit a step-like behavior caused by the conductance peak in the dot.
(e)~An accurate dc measurement (precision $10^{-4}$) reveals the step in $G_\mathrm{ring}^\mathrm{dc}(V_\mathrm{pg}^\mathrm{dc})$. The global evolution
     of $G_\mathrm{ring}^\mathrm{dc}$ differs from the estimated value in (d) because of a major charge rearrangement between the two measurements. Measurements were performed at $B = 4$\,mT on an AB interference minimum. 
}\label{fig2}\end{center}\end{figure}

In contrast to the dc-AB amplitude $G_1^\mathrm{dc}$, the amplitude of the AB oscillations in the transconductance $G_1^\mathrm{trans}$ depends strongly on the dc  dot conductance $G_\mathrm{dot}$ measured simultaneously. Figure \,\ref{fig2}(a) shows the results for two settings of the plunger-gate voltage $V_\mathrm{pg}^\mathrm{dc}$ [indicated by the bold and dashed vertical lines in Fig.\,\ref{fig2}(b)], leading to different amplitudes in $G_1^\mathrm{trans}$. The amplitude $G_1^\mathrm{trans}$ is obtained from a series of such measurements in the magnetic field range shown in Fig.\,\ref{fig2}(a) and plotted versus plunger-gate voltage together with the corresponding quantum dot conductance in Fig.\,\ref{fig2}(b). The AB amplitude $G_1^\mathrm{trans}$ through the ring shows a dip exactly at the gate voltage where the dc conductance through the quantum dot $G_\mathrm{dot}$ displays a conductance maximum. 

This dip is found for a series of Coulomb peaks and we verified that each dip shifts in plunger-gate voltage together with the corresponding conductance peak in $G_\mathrm{dot}^\mathrm{dc}$ if the latter is moved by a remote gate electrode. This proves the intimate relation between the two effects. The absolute magnitude of the dip in the transconductance depends on how large the transconductance is. For regions, where the AB amplitude $G_1^\mathrm{trans}$ is close to zero, i.e., where the dc signal of the ring $G_1^\mathrm{dc}$ shows a weak dependence on the gate voltage $V_\mathrm{pg}^\mathrm{dc}$ (close to a maximum or a minimum), we cannot detect a dip. For negative  values of $G_1^\mathrm{trans}$ the dip effectively becomes a peak as expected (not shown).

An equivalent result to Fig.\,\ref{fig2}(b) is presented in Fig.\,\ref{fig2}(c) for $G_\mathrm{ring}^\mathrm{trans}(B)$ measured at fixed magnetic field $B=4$\,mT, where the effect of the dip in $G_1^\mathrm{trans}$ is strongest [black vertical line in Fig.\, \ref{fig2}(a)].
Both, $h/e$- and $h/2e$-periodic oscillations can be detected in the ac and the dc-AB measurements. The amplitude of the $h/2e$ effect is generally much weaker and sometimes vanishes in the noise. In parameter regimes where both effects are detectable, they behave very similarly, i.e., also the amplitude of the ac $h/2e$-oscillations shows a dip where the dot conductance displays a maximum. A gate voltage applied on the plunger depletes slightly the top arm of the ring, therefore introducing an asymmetry in the transmissions along the two paths of the interferometer. This is seen as a reduction of the contrast of the oscillations.

\begin{figure}[t]
\begin{center}
\includegraphics[width=3.4in]{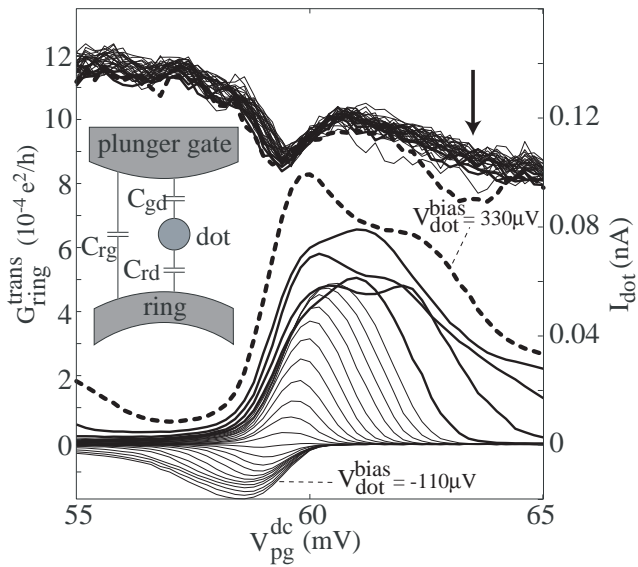}
\caption{Dot current and ring transconductance for a dot bias from $V_{\mathrm dot}^{\mathrm bias} = -110\mu V$ to $V_{\mathrm dot}^{\mathrm bias} = +110\mu V$ (thin lines). A
    higher dot bias voltage increases the current through the dot. For small bias
    voltages, the dot I-V characteristic is linear (on a conductance peak). At higher bias voltages  $V_{\mathrm dot}^{\mathrm bias} = 130\mu V,180\mu V,230\mu V,280\mu V,330\mu V$ (bold and dashed line), new dot levels become available for transport. The dip in the transconductance $G_\mathrm{ring}^\mathrm{trans}$ measured at fixed
    magnetic field does not change, except for the highest bias voltage (dashed line), where
    a second dip appears (see vertical arrow). Inset: Electrostatic model for the coupling of the dot to the ring. The notation is described in the text.}\label{fig3}\end{center}\end{figure}

Our interpretation of the reduced AB amplitude is based on screening of the modulated plunger-gate voltage felt in the ring due to single-electron charging in the dot on a conductance peak.
The screening effect can be understood in a capacitive model (inset of Fig.\,\ref{fig3}). There is a direct capacitive coupling $C_\mathrm{rg}$ between plunger gate and ring arm. In addition, the capacitances between gate and dot, $C_\mathrm{gd}$, and dot and ring, $C_\mathrm{rd}$, are connected in series, parallel to $C_\mathrm{rg}$. The induced charge on the arm of the ring is given by
\[ Q_\mathrm{ring}=\underbrace{\left[\left(\frac{1}{C_\mathrm{rd}}+\frac{1}{C_\mathrm{gd}}\right)^{-1}+C_\mathrm{rg}\right]}_{C_x}V_\mathrm{pg}^\mathrm{dc}+\underbrace{\frac{C_\mathrm{rd}}{C_\mathrm{dd}}}_{\gamma}Q_\mathrm{dot}, \]  

where $C_\mathrm{dd}=-(C_{rd}+C_{gd})$ is the self-capacitance of the dot. 
When the charge on the dot, $Q_\mathrm{dot}$, changes by $-\left| e\right|$ on a conductance peak, $Q_\mathrm{ring}$ changes by a small amount $\Delta Q_\mathrm{ring}$. By sweeping $V_\mathrm{pg}^\mathrm{dc}$ over a conductance peak in the dot, the thermal smearing of the peak will also smear the step in $Q_\mathrm{ring}$. 
We then assume that the corresponding local potential change in the ring is $\Delta U_\mathrm{ring}\propto \Delta Q_\mathrm{ring}$ and a step in $\Delta U_\mathrm{ring}$ will lead to a step in all the conductances $G_i^\mathrm{dc}$ as a function of $V_\mathrm{pg}^\mathrm{dc}$. Since the transconductance measures the derivative of the ring current for a fixed bias voltage $V_\mathrm{ring}^\mathrm{bias}$, the step-like behavior in $G_\mathrm{ring}^\mathrm{dc}$ appears as a dip in $G_\mathrm{ring}^\mathrm{trans}$ at a value of $V_\mathrm{pg}^\mathrm{dc}$ where a conductance peak occurs in the dot. 

In order to check this electrostatic model we have integrated the transconductance $G_\mathrm{ring}^\mathrm{trans}$ presented in Fig. \,\ref{fig2}(c) over the gate voltage, i.e., $G_\mathrm{ring}^\mathrm{int}=\int(G_\mathrm{ring}^\mathrm{trans}/V_\mathrm{ring}^\mathrm{bias}) dV_\mathrm{pg}^\mathrm{dc}$ and found a small kink as presented in Fig.\,\ref{fig2}(d). For a consistency check we have measured the behavior of the dc-AB effect directly with great accuracy, which requires measurement times of the order of days for one trace. The result $G_\mathrm{ring}^\mathrm{dc}$ [Fig.\,\ref{fig2} (e)], which shows effects with an accuracy better than $10^{-3}$, is very similar to the integrated data. The position of the step is slightly different between the two measurements due to a charge rearrangement which required us to readjust the gates slightly.

Assuming the ring conductance to be a smooth function of the induced charge $Q_\mathrm{ring}$, the factor $C_x$ relates to the slope of the ring conductance as a function of gate voltage in Fig.\,\ref{fig2}(d-e) while $\gamma$ is related to the sensitivity of the ring on the charge state of the dot, i.e. the size of the step in the ring conductance. From Fig.\,\ref{fig2}(d-e) we estimate the change in gate voltage $\Delta V_{\mathrm{pg}}^{\mathrm{dc}}=-|e|\gamma/C_x\approx0.4\,mV$ which is necessary to compensate for the addition of one electron to the dot. 

The change in the local ring potential $\Delta U_\mathrm{ring}$ due to an additional electron on the dot can be estimated using the geometric ring--dot separation and taking the image charge in the top gate into account\;\cite{03ihn}. The resulting energy scale of 8\,$\mu$V is also consistent with temperature-dependent measurements: the dip in $G_1^\mathrm{trans}$ weakens with increasing temperature and disappears above $T=190$\,mK.

From measuring, on the one hand, the lever arms between the plunger gate and the dot, and, on the other hand, between the plunger gate and a remote dot located on the other side of the ring, we estimate the plunger gate--ring lever arm by interpolation to be 0.02. A modulation $V_\mathrm{pg}^\mathrm{ac}=300 \mu$V corresponds to a potential modulation of about 6 $\mu$eV in the ring. 

In our model the kinks in the dc conductance are similar in origin to the effects observed in Ref.\;\cite{93field}, where kinks are found in the conductance of a quantum point contact neighboring a quantum dot. However, our results are different in that we find this effect also in the phase-coherent part of the detector signal, namely, the AB amplitude $G_1^\mathrm{dc}$. This raises the question if the strong suppression of the AB oscillation amplitude on a conductance peak could also be related to dephasing of partial waves in the ring due to the fluctuating charge in the dot on a conductance peak. In the experiment of Buks et al.\;\cite{98buks} it was found that the dc AB amplitude showed a reduction when the current through a nearby quantum point contact was increased, which in our case would correspond to a reduction of the AB-effect if the dot current is increased.

Figure\,\ref{fig3} shows the dot current $I_\mathrm{dot}$ together with $G_\mathrm{ring}^\mathrm{trans}$ as a function of dot bias. The position and strength of the dip in $G_\mathrm{ring}^\mathrm{trans}$ is basically bias-independent. The dc dot-conductance maximum grows with increasing bias and shifts in position. This follows from an examination of the Coulomb diamonds which show a pronounced asymmetry for positive and negative biases, in particular, for excited states that become important for large biases (bold lines). For the largest dot bias of $330\,\mu$V (indicated by a thick dashed line), a weak second dip occurs in the transconductance around $V_\mathrm{pg}^\mathrm{dc} \approx 62$\,mV (see vertical arrow in Fig.\,\ref{fig3}). For about the same gate voltage we find a shoulder in the dot conductance and in most cases we could link such a behavior to the  excited state spectrum. 
If dephasing were the dominant meachnism, the dip depth should strongly increase with the current level through the dot, which is clearly not the case in our experiment.

This can be understood from the fact that the current through a constriction\;\cite{98buks} can be much larger than that through a quantum dot which has to be kept in the Coulomb-blockade regime. In our case the current through the ring is of the order of nanoamperes, while the dot current is in the range of picoamperes. The dwell time of an electron in the ring is considerably shorter than the dwell time of an electron in the dot. As a consequence, the ring is most of the time unobserved. For an electron passing through the ring, the probability to be dephased by scattering with an electron passing through the dot is therefore rather small. 
Thus we conclude that such an interpretation is not significant for the observed AB amplitude reduction and the electrostatic model seems to be perfectly valid also for the phase-coherent signal of a quantum ring detector. 


We have demonstrated that the phase-coherent AB oscillations in a quantum ring are sensitive to single-electron charging of an adjacent quantum dot. The ring can be used as a phase-coherent detector of the charge state of the dot.

We are grateful to B. Altshuler for useful discussions. Financial support from the Swiss Science Foundation (Schweizerischer Nationalfonds) is acknowledged.

\bibliographystyle{apsrev}

\begin{thebibliography}{20}
\expandafter\ifx\csname natexlab\endcsname\relax\def\natexlab#1{#1}\fi
\expandafter\ifx\csname bibnamefont\endcsname\relax
  \def\bibnamefont#1{#1}\fi
\expandafter\ifx\csname bibfnamefont\endcsname\relax
  \def\bibfnamefont#1{#1}\fi
\expandafter\ifx\csname citenamefont\endcsname\relax
  \def\citenamefont#1{#1}\fi
\expandafter\ifx\csname url\endcsname\relax
  \def\url#1{\texttt{#1}}\fi
\expandafter\ifx\csname urlprefix\endcsname\relax\def\urlprefix{URL }\fi
\providecommand{\bibinfo}[2]{#2}
\providecommand{\eprint}[2][]{\url{#2}}

\bibitem[{\citenamefont{Levinson}(1997)}]{97levinson}
\bibinfo{author}{\bibfnamefont{Y.}~\bibnamefont{Levinson}},
  \bibinfo{journal}{Europhys. Lett.} \textbf{\bibinfo{volume}{39}},
  \bibinfo{pages}{299} (\bibinfo{year}{1997}).

\bibitem[{\citenamefont{Gurvitz}(1997)}]{97gurvitz}
\bibinfo{author}{\bibfnamefont{S.~A.} \bibnamefont{Gurvitz}},
  \bibinfo{journal}{Phys. Rev. B} p. \bibinfo{pages}{15215}
  (\bibinfo{year}{1997}).

\bibitem[{\citenamefont{B{\"u}ttiker and Martin}(2000)}]{00buttiker}
\bibinfo{author}{\bibfnamefont{M.}~\bibnamefont{B{\"u}ttiker}}
  \bibnamefont{and} \bibinfo{author}{\bibfnamefont{A.}~\bibnamefont{Martin}},
  \bibinfo{journal}{Phys. Rev. B} \textbf{\bibinfo{volume}{61}},
  \bibinfo{pages}{2737} (\bibinfo{year}{2000}).

\bibitem[{\citenamefont{Pohjola et~al.}(2001)\citenamefont{Pohjola, Schoeller,
  and Sch{\"o}n}}]{01pohjola}
\bibinfo{author}{\bibfnamefont{T.}~\bibnamefont{Pohjola}},
  \bibinfo{author}{\bibfnamefont{H.}~\bibnamefont{Schoeller}},
  \bibnamefont{and}
  \bibinfo{author}{\bibfnamefont{G.}~\bibnamefont{Sch{\"o}n}},
  \bibinfo{journal}{Europhys. Lett.} \textbf{\bibinfo{volume}{54}},
  \bibinfo{pages}{241} (\bibinfo{year}{2001}).

\bibitem[{\citenamefont{S.Pilgram and M.B{\"u}ttiker}(2002)}]{02pilgram}
\bibinfo{author}{\bibnamefont{S.Pilgram}} \bibnamefont{and}
  \bibinfo{author}{\bibnamefont{M.B{\"u}ttiker}}, \bibinfo{journal}{Phys. Rev.
  Lett.} \textbf{\bibinfo{volume}{89}}, \bibinfo{pages}{200401}
  (\bibinfo{year}{2002}).

\bibitem[{\citenamefont{Field et~al.}(1993)\citenamefont{Field, Smith, Pepper,
  Richie, Frost, Jones, and Hasko}}]{93field}
\bibinfo{author}{\bibfnamefont{M.}~\bibnamefont{Field}},
  \bibinfo{author}{\bibfnamefont{C.~G.} \bibnamefont{Smith}},
  \bibinfo{author}{\bibfnamefont{M.}~\bibnamefont{Pepper}},
  \bibinfo{author}{\bibfnamefont{D.~A.} \bibnamefont{Richie}},
  \bibinfo{author}{\bibfnamefont{J.~E.~F.} \bibnamefont{Frost}},
  \bibinfo{author}{\bibfnamefont{G.~A.~C.} \bibnamefont{Jones}},
  \bibnamefont{and} \bibinfo{author}{\bibfnamefont{D.~G.} \bibnamefont{Hasko}},
  \bibinfo{journal}{Phys. Rev. Lett} \textbf{\bibinfo{volume}{70}},
  \bibinfo{pages}{1311} (\bibinfo{year}{1993}).

\bibitem[{\citenamefont{Cooper et~al.}(2000)\citenamefont{Cooper, Smith,
  Ritchie, Linfield, Jin, and Launois}}]{00cooper}
\bibinfo{author}{\bibfnamefont{J.}~\bibnamefont{Cooper}},
  \bibinfo{author}{\bibfnamefont{C.~G.} \bibnamefont{Smith}},
  \bibinfo{author}{\bibfnamefont{D.~A.} \bibnamefont{Ritchie}},
  \bibinfo{author}{\bibfnamefont{E.~H.} \bibnamefont{Linfield}},
  \bibinfo{author}{\bibfnamefont{Y.}~\bibnamefont{Jin}}, \bibnamefont{and}
  \bibinfo{author}{\bibfnamefont{H.}~\bibnamefont{Launois}},
  \bibinfo{journal}{Physica E} \textbf{\bibinfo{volume}{6}},
  \bibinfo{pages}{457} (\bibinfo{year}{2000}).

\bibitem[{\citenamefont{Sprinzak et~al.}(2002)\citenamefont{Sprinzak, Ji,
  Heiblum, Mahalu, and Shtrikman}}]{02sprinzak}
\bibinfo{author}{\bibfnamefont{D.}~\bibnamefont{Sprinzak}},
  \bibinfo{author}{\bibfnamefont{Y.}~\bibnamefont{Ji}},
  \bibinfo{author}{\bibfnamefont{M.}~\bibnamefont{Heiblum}},
  \bibinfo{author}{\bibfnamefont{D.}~\bibnamefont{Mahalu}}, \bibnamefont{and}
  \bibinfo{author}{\bibfnamefont{H.}~\bibnamefont{Shtrikman}},
  \bibinfo{journal}{Phys. Rev. Lett} \textbf{\bibinfo{volume}{88}},
  \bibinfo{pages}{176805} (\bibinfo{year}{2002}).

\bibitem[{\citenamefont{Smith et~al.}(2002)\citenamefont{Smith, Gardelis,
  Cooper, Ritchie, Linfield, Jin, and Launois}}]{02smith}
\bibinfo{author}{\bibfnamefont{C.~G.} \bibnamefont{Smith}},
  \bibinfo{author}{\bibfnamefont{S.}~\bibnamefont{Gardelis}},
  \bibinfo{author}{\bibfnamefont{J.}~\bibnamefont{Cooper}},
  \bibinfo{author}{\bibfnamefont{D.~A.} \bibnamefont{Ritchie}},
  \bibinfo{author}{\bibfnamefont{E.~H.} \bibnamefont{Linfield}},
  \bibinfo{author}{\bibfnamefont{Y.}~\bibnamefont{Jin}}, \bibnamefont{and}
  \bibinfo{author}{\bibfnamefont{H.}~\bibnamefont{Launois}},
  \bibinfo{journal}{Physica E} \textbf{\bibinfo{volume}{12}},
  \bibinfo{pages}{830} (\bibinfo{year}{2002}).

\bibitem[{\citenamefont{Gardelis et~al.}(2003)\citenamefont{Gardelis, Smith,
  Cooper, Ritchie, Linefield, Jin, and Launois}}]{03gardelis}
\bibinfo{author}{\bibfnamefont{S.}~\bibnamefont{Gardelis}},
  \bibinfo{author}{\bibfnamefont{C.}~\bibnamefont{Smith}},
  \bibinfo{author}{\bibfnamefont{J.}~\bibnamefont{Cooper}},
  \bibinfo{author}{\bibfnamefont{D.}~\bibnamefont{Ritchie}},
  \bibinfo{author}{\bibfnamefont{E.}~\bibnamefont{Linefield}},
  \bibinfo{author}{\bibfnamefont{Y.}~\bibnamefont{Jin}}, \bibnamefont{and}
  \bibinfo{author}{\bibfnamefont{H.}~\bibnamefont{Launois}},
  \bibinfo{journal}{Phys. Rev. B} \textbf{\bibinfo{volume}{67}},
  \bibinfo{pages}{073302} (\bibinfo{year}{2003}).

\bibitem[{\citenamefont{Elzerman et~al.}(2003)\citenamefont{Elzerman, Hanson,
  Greidanus, van Beveren, Franceschi, Vandersypen, Tarucha, and
  Kouwenhoven}}]{03elzerman}
\bibinfo{author}{\bibfnamefont{J.~M.} \bibnamefont{Elzerman}},
  \bibinfo{author}{\bibfnamefont{R.}~\bibnamefont{Hanson}},
  \bibinfo{author}{\bibfnamefont{J.~S.} \bibnamefont{Greidanus}},
  \bibinfo{author}{\bibfnamefont{L.~H.~W.} \bibnamefont{van Beveren}},
  \bibinfo{author}{\bibfnamefont{S.~D.} \bibnamefont{Franceschi}},
  \bibinfo{author}{\bibfnamefont{L.~M.~K.} \bibnamefont{Vandersypen}},
  \bibinfo{author}{\bibfnamefont{S.}~\bibnamefont{Tarucha}}, \bibnamefont{and}
  \bibinfo{author}{\bibfnamefont{L.~P.} \bibnamefont{Kouwenhoven}},
  \bibinfo{journal}{Phys. Rev. B} \textbf{\bibinfo{volume}{67}},
  \bibinfo{pages}{161308(R)} (\bibinfo{year}{2003}).

\bibitem[{\citenamefont{Kataoka et~al.}(2000)\citenamefont{Kataoka, Ford,
  Faini, Mailly, Simmons, Mace, Liang, and Ritchie}}]{00kataoka}
\bibinfo{author}{\bibfnamefont{M.}~\bibnamefont{Kataoka}},
  \bibinfo{author}{\bibfnamefont{C.~J.~B.} \bibnamefont{Ford}},
  \bibinfo{author}{\bibfnamefont{G.}~\bibnamefont{Faini}},
  \bibinfo{author}{\bibfnamefont{D.}~\bibnamefont{Mailly}},
  \bibinfo{author}{\bibfnamefont{M.~Y.} \bibnamefont{Simmons}},
  \bibinfo{author}{\bibfnamefont{D.~R.} \bibnamefont{Mace}},
  \bibinfo{author}{\bibfnamefont{C.~T.} \bibnamefont{Liang}}, \bibnamefont{and}
  \bibinfo{author}{\bibfnamefont{D.~A.} \bibnamefont{Ritchie}},
  \bibinfo{journal}{Physica E} \textbf{\bibinfo{volume}{6}},
  \bibinfo{pages}{495} (\bibinfo{year}{2000}).

\bibitem[{\citenamefont{Wilhelm et~al.}(2002)\citenamefont{Wilhelm, Schmid,
  Weis, and v.~Klitzing}}]{02wilhelm}
\bibinfo{author}{\bibfnamefont{U.}~\bibnamefont{Wilhelm}},
  \bibinfo{author}{\bibfnamefont{J.}~\bibnamefont{Schmid}},
  \bibinfo{author}{\bibfnamefont{J.}~\bibnamefont{Weis}}, \bibnamefont{and}
  \bibinfo{author}{\bibfnamefont{K.}~\bibnamefont{v.~Klitzing}},
  \bibinfo{journal}{Physica E} \textbf{\bibinfo{volume}{14}},
  \bibinfo{pages}{385} (\bibinfo{year}{2002}).

\bibitem[{\citenamefont{Aharonov and Bohm}(1959)}]{59aharonov}
\bibinfo{author}{\bibfnamefont{Y.}~\bibnamefont{Aharonov}} \bibnamefont{and}
  \bibinfo{author}{\bibfnamefont{D.}~\bibnamefont{Bohm}},
  \bibinfo{journal}{Phys. Rev.} \textbf{\bibinfo{volume}{115}},
  \bibinfo{pages}{485} (\bibinfo{year}{1959}).

\bibitem[{\citenamefont{Timp et~al.}(1987)\citenamefont{Timp, Chang,
  Cunningham, Chang, Mankiewich, Behringer, and Howard}}]{87timp}
\bibinfo{author}{\bibfnamefont{G.}~\bibnamefont{Timp}},
  \bibinfo{author}{\bibfnamefont{A.~M.} \bibnamefont{Chang}},
  \bibinfo{author}{\bibfnamefont{J.~E.} \bibnamefont{Cunningham}},
  \bibinfo{author}{\bibfnamefont{T.~Y.} \bibnamefont{Chang}},
  \bibinfo{author}{\bibfnamefont{P.}~\bibnamefont{Mankiewich}},
  \bibinfo{author}{\bibfnamefont{R.}~\bibnamefont{Behringer}},
  \bibnamefont{and} \bibinfo{author}{\bibfnamefont{R.~E.}
  \bibnamefont{Howard}}, \bibinfo{journal}{Phys. Rev. Lett}
  \textbf{\bibinfo{volume}{58}}, \bibinfo{pages}{2814} (\bibinfo{year}{1987}).

\bibitem[{\citenamefont{Hansen et~al.}(2001)\citenamefont{Hansen, Kristensen,
  Pedersen, So{\^i}rensen, and Lindelof}}]{01hansen}
\bibinfo{author}{\bibfnamefont{A.~E.} \bibnamefont{Hansen}},
  \bibinfo{author}{\bibfnamefont{A.}~\bibnamefont{Kristensen}},
  \bibinfo{author}{\bibfnamefont{S.}~\bibnamefont{Pedersen}},
  \bibinfo{author}{\bibfnamefont{C.~B.} \bibnamefont{So{\^i}rensen}},
  \bibnamefont{and} \bibinfo{author}{\bibfnamefont{P.~E.}
  \bibnamefont{Lindelof}}, \bibinfo{journal}{Phys. Rev. B}
  \textbf{\bibinfo{volume}{64}}, \bibinfo{pages}{045327}
  (\bibinfo{year}{2001}).

\bibitem[{\citenamefont{Ihn et~al.}(2003)\citenamefont{Ihn, Fuhrer, M.Sigrist,
  Ensslin, Wegscheider, and Bichler}}]{03ihn}
\bibinfo{author}{\bibfnamefont{T.}~\bibnamefont{Ihn}},
  \bibinfo{author}{\bibfnamefont{A.}~\bibnamefont{Fuhrer}},
  \bibinfo{author}{\bibnamefont{M.Sigrist}},
  \bibinfo{author}{\bibfnamefont{K.}~\bibnamefont{Ensslin}},
  \bibinfo{author}{\bibfnamefont{W.}~\bibnamefont{Wegscheider}},
  \bibnamefont{and} \bibinfo{author}{\bibfnamefont{M.}~\bibnamefont{Bichler}},
  \bibinfo{journal}{Adv. In Solid State Phys.} \textbf{\bibinfo{volume}{43}},
  \bibinfo{pages}{139} (\bibinfo{year}{2003}).

\bibitem[{\citenamefont{Buks et~al.}(1998)\citenamefont{Buks, Schuster,
  Heiblum, Mahalu, and Umansky}}]{98buks}
\bibinfo{author}{\bibfnamefont{E.}~\bibnamefont{Buks}},
  \bibinfo{author}{\bibfnamefont{R.}~\bibnamefont{Schuster}},
  \bibinfo{author}{\bibfnamefont{M.}~\bibnamefont{Heiblum}},
  \bibinfo{author}{\bibfnamefont{D.}~\bibnamefont{Mahalu}}, \bibnamefont{and}
  \bibinfo{author}{\bibfnamefont{V.}~\bibnamefont{Umansky}},
  \bibinfo{journal}{Nature} \textbf{\bibinfo{volume}{391}},
  \bibinfo{pages}{871} (\bibinfo{year}{1998}).

\bibitem[{\citenamefont{Fuhrer et~al.}(2002)\citenamefont{Fuhrer, Dorn,
  L{\"u}scher, Heinzel, Ensslin, Wegscheider, and Bichler}}]{02fuhrer}
\bibinfo{author}{\bibfnamefont{A.}~\bibnamefont{Fuhrer}},
  \bibinfo{author}{\bibfnamefont{A.}~\bibnamefont{Dorn}},
  \bibinfo{author}{\bibfnamefont{S.}~\bibnamefont{L{\"u}scher}},
  \bibinfo{author}{\bibfnamefont{T.}~\bibnamefont{Heinzel}},
  \bibinfo{author}{\bibfnamefont{K.}~\bibnamefont{Ensslin}},
  \bibinfo{author}{\bibfnamefont{W.}~\bibnamefont{Wegscheider}},
  \bibnamefont{and} \bibinfo{author}{\bibfnamefont{M.}~\bibnamefont{Bichler}},
  \bibinfo{journal}{Superlatice Microst.} \textbf{\bibinfo{volume}{31}},
  \bibinfo{pages}{19} (\bibinfo{year}{2002}).

\bibitem[{\citenamefont{Ihn}(2004)}]{04ihn}
\bibinfo{author}{\bibfnamefont{T.}~\bibnamefont{Ihn}},
  \emph{\bibinfo{title}{Electronic Quantum Transport in Mesoscopic
  Semiconductor Structures}}, vol. \bibinfo{volume}{192} of
  \emph{\bibinfo{series}{Springer Tracts in Modern Physics}}
  (\bibinfo{publisher}{Springer}, \bibinfo{address}{New York},
  \bibinfo{year}{2004}).

\end{thebibliography}


\end{document}